\documentclass[aps,pra,twocolumn,floatfix,showpacs]{revtex4-1}
\usepackage{amsmath}
\usepackage{amsfonts}
\usepackage{amssymb}
\usepackage{graphicx}
\usepackage{color,soul}

\begin{document}

\title{Optomechanical Ramsey Interferometry}
\author{Kenan Qu$^{1}$}
\author{Chunhua Dong$^{2,3}$}
\email{chunhua@ustc.edu.cn}
\author{Hailin Wang$^{4}$}
\author{G. S. Agarwal$^{1}$}
\affiliation{$^{1}$Department of Physics, Oklahoma State University, Stillwater,
Oklahoma, 74078, USA}
\affiliation{$^{2}$Key Laboratory of Quantum Information, University of Science and
Technology of China, Hefei 230026, P. R. China}
\affiliation{$^{3}$Synergetic Innovation Center of Quantum Information and Quantum
Physics, University of Science and Technology of China, Hefei, Anhui 230026,
P. R. China.}
\affiliation{$^{4}$Department of Physics, University of Oregon, Eugene, Oregon 97403, USA}
\date{\today }

\begin{abstract}
We adopt the Ramsey's method of separated oscillatory fields to study
coherences of the mechanical system in an optomechanical resonator. The high resolution
Ramsey fringes are observed in the emission optical field, when two pulses
separated in time are applied. We develop a theory to describe the transient
optomechanical behavior underlying the Ramsey fringes. We also perform the
experimental demonstration using a silica microresonator. The method is
versatile and can be adopted to different types of mechanical resonators,
electromechanical resonators.
\end{abstract}

\maketitle

\section{Introduction}
The Ramsey method~\cite{Ramsey0,Ramsey1,Ramsey2,Ramsey3} of separated
oscillatory fields is a highly successful method of precision spectroscopy
and has been extensively used in a wide spectral range starting from radio
frequency to optical domain. This method has yielded the atomic and
molecular transition frequencies with very high precision especially by
using phase coherent pulses with a duration short compared to the atomic
decay times. The Ramsey technique is an interference techniques in which one
studies the result of the quantum mechanical amplitudes in different domains
where fields are applied. The Ramsey's interferometric technique has so far
been used in the context of the study of the phase coherence in atomic and
molecular systems~\cite{atom1,atom2,atom3}. Ramsey method has been
especially successful in the detection of quantum coherences, such as in the
detection of the Cat states of the electromagnetic field~\cite{Haroche1,Haroche2,Haroche3}. In this paper, we present a demonstration of the Ramsey interferometry (RI) in the context of a
macroscopic system like a nanomechanical system. The Ramsey interferometry
is especially important in probing the dynamics of the nanomechanical system
as the detected interference signals would be sensitive to any dynamical
changes in the mechanical oscillator in the time between which the pulses
are off. The dynamics is a direct measure of the coherence time of the mechanical mirror which is an important issue in precision measurements~\cite{review1}.
We note that a variety of coherent techniques have been used to
study optomechanical and electromechanical systems. These techniques
primarily make use of the electromagnetically induced transparency and
absorption~\cite{EIT1,EIT2,EIT3,EIA,slowlight} and the resulting
applications like storage of light~\cite{Dong}, slow light~\cite{slowlight},
etc. We also note that in a recent work~\cite{Faust}, Ramsey interferometry is used to probe the dynamical behavior of two nanostring resonators coupled by an rf field. These authors realized a two level system in this manner and then applied the technique of Ramsey interferometry as originally developed by Ramsey. In our work we perform Ramsey interferometry using an optical mode and a mechanical mode.

In atomic systems, the RI uses two spatially or temporally separated laser
pulses to interact with atoms and detects the excitation probability of the
atoms. Each laser pulse drives the atomic transition between the ground
state $|g\rangle $ and excited state $|e\rangle $. The atomic transition
probability after a single weak pulse excitation~\cite{Ramsey1} has a form~\cite{Ramsey1} of $p_{s}=g^{2}\tau^{2}\mathrm{sinc}^{2}(\Delta \tau/2)$, where $\Delta $ is the frequency
detuning between the laser and the atomic transition frequency. For simplicity, we use square pulses although other pulse forms can be used.  Ramsey's two-pulse scheme yields sharper interference patterns without requiring a long interaction time. It has the excitation
probability $p_{R}=p_{s}(\cos \Delta T+1)$. Here the pulse separation in
time $T$ is shorter than the decay time of the atoms. Note that we use weak pulses in contrast to the case where Rabi oscillations are used. The pattern arises
from the quantum interference of two paths in which the atom can get
excited. The RI is expected to be useful whenever we have a system with a
long decay time. Therefore, we make use of the very long decay time of the
mechanical coherence~\cite{Dong,highQ,disk} to demonstrate Ramsey
interferometry in an optomechanical system (OMS).

There are similarities between this work and the storage work~\cite{Dong}. In storage work, one demonstrates the ability of the optomechanical system to store light pulses and to recover these. In RI, the objective is quite different: Here one wants to have a sensitive probe of the dynamics of the mirror and even a sensitive probe of the displacements. Thus in storage, the focus is on optical pulses whereas in RI the focus is on the mechanical element.

The organization of this paper is as follows: In Sec. II, we describe in detail the experimental setup and method of RI in optomechanics; we also show the results exhibiting Ramsey interference patterns. In Sec. III, we theoretically explain the phenomenon starting from the Hamiltonian and obtain the condition to see Ramsey pattern in optomechanics. In Sec. VI we present our conclusions. Since our experiment is done with fields in the classical domain, we obtain theoretical results valid in this regime only.

\section{Experiment and result}
\begin{figure}[tbph]
\includegraphics[width=0.48\textwidth]{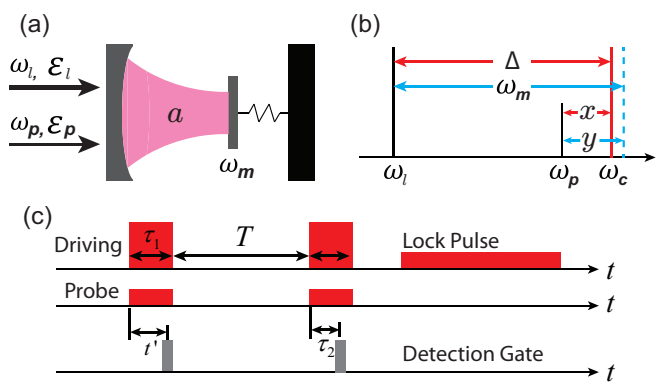}
\caption{{\ (a) Sketch of the OMS. (b) The frequency relations of different
modes in OMS. The labellings are explained in the text. (c) Sketch of the
Ramsey pulses sequence applied to OMS.}}
\label{fig1}
\end{figure}

In this article, we consider a generic OMS which contains a Fabry-Perot
optical cavity with an oscillating mirror on one side, as in Fig.~\ref{fig1}%
a. The motion of this mirror is driven by the radiation pressure force of
the optical mode. This model is applicable to a large number of other
systems, such as electromechanical systems \cite{EMS1}, Brillouin modes in
microfluidic devices \cite{EMS2}, and mechanical breathing modes in silica
microspheres \cite{Dong}, microtoroids \cite{highQ}, or microdisks \cite%
{disk}. For the proposed optomechanical RI, we enable the optomechanical
coupling in two time-separated regions, during which a pair of laser pulses
including both driving pulse and probe pulse are sent into the cavity (see
Fig.~\ref{fig1}a). The probe laser, with frequency $\omega _{p}$, is near
the cavity resonance, $\omega _{c}$ and the driving laser, with frequency $\omega _{l}$, is near the red sideband of the cavity resonance, $\omega_{c}-\omega _{m}$, with $\omega _{m}$ being the mechanical frequency (see
Fig.\ref{fig1}b). For convenience, we also define two frequency parameters $x=\Delta-\omega_p+\omega_l$, $y=\omega_m-\omega_p+\omega_l$. Note that $x\approx \omega _{c}-\omega _{p}$ and is close to zero if the probe field is on resonance with the cavity frequency. Then $y\sim 0$ if $\omega
_{l}\sim \omega _{c}-\omega _{m}$.  The pulse sequence is shown in Fig.~\ref{fig1}(c), where
we denote the widths of the pulses by $\tau_{1}$ and $\tau_{2}$ and the
separation by T. The time-dependent amplitudes of the driving and probe, $%
\mathcal{E}_{l}(t)$ and $\mathcal{E}_{p}(t)$, are both nonzero only during
the pulses: $\mathcal{E}_{l}(t)=\mathcal{E}_{l}$ and $\mathcal{E}_{p}(t)=%
\mathcal{E}_{p}$ for $t\in \lbrack 0,\tau_{1}]\cup \lbrack T+\tau
_{1},T+\tau_{1}+\tau_{2}]$, and $\mathcal{E}_{l}(t)=\mathcal{E}_{p}(t)=0$
otherwise.

\begin{figure}[!bph]
\includegraphics[width=0.48\textwidth]{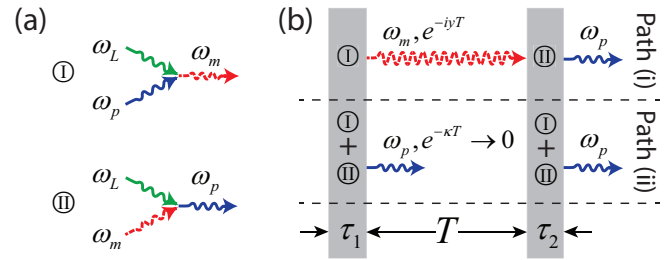}
\caption{{The physical process of the photons and phonons in the Ramsey
setup. The solid arrows refer to photons and dashed ones refer to phonons.
(a) the driving photon combining a probe photon generates a phonon and the
driving photon combining a photon generates a probe photon. (b) the two
paths of generating a photon at $\protect\omega _{p}$ and $y=\omega_l+\omega_m-\omega_p$. }}
\label{fig2}
\end{figure}

In a RI setup, two pairs of pulses with separation T are sent to the cavity.
Two processes are taking place when a pulse pair is in the cavity: %
\textcircled{\scriptsize  I} the coupling and probe photons combine and
produce coherent phonons; and \textcircled{\scriptsize  II} the coherent
phonons combine with the coupling photons and generate an anti-Stokes
sideband near the cavity resonance. The application of the first pulse pair
creates both coherent phonons and cavity photons. After the first pulse
pair, the optical mode decays rapidly during the free evolution and it
becomes negligible as $\mathrm{e}^{-(\kappa /2)T}\thicksim 0$, where $\kappa$ is the total decay rate of the cavity mode. On the other than,  the mechanical mode shows almost no decay as $\mathrm{e}^{-(\gamma _{m}/2)T}\approx1$, where $\gamma _{m}$ is the mechanical damping rate. This is because $\gamma_m\ll\kappa$.  Thus, before the second pulse pair is applied, the mechanical mode barely decays but gathers a phase
$\omega_m T$.  Now
we examine the two paths which lead to the interference in the optical field
produced at $\omega _{p}$. The phonon created in the zone \textquotedblleft $%
\tau_{1}$\textquotedblright\ survives and interacts with the driving laser
to produce a photon at $\omega _{p}$ via process
\textcircled{\scriptsize II} in the zone \textquotedblleft $\tau_{2}$\textquotedblright . This is marked as path (i) in Fig.~\ref{fig2}. Photons at
$\omega _{p}$ can also be generated entirely in the zone \textquotedblleft $%
\tau_{2}$\textquotedblright , as discussed earlier [path (ii) of Fig.~\ref{fig2}]. These two paths are
displayed in Fig.~\ref{fig2}b and their coherent character leads to Ramsey
fringes in the optical output field, which can be detected directly through
heterodyne interference~\cite{GSAbook} with a local oscillator.
Note that the pattern does not arise from the direct interference of the two input probe pulses, since the the free evolution time is much longer than the optical decay time, $T\gg1/\kappa$. The mechanical oscillation is the only medium that can carry coherence during both pulses. Therefore, the fringes arise from the mechanical coherence effects although we observe such coherences in the optical fields. In the optomechanical RI, we take advantage of the long life time of the phonons in the mechanical resonator, as demonstrated in the previous optical pulse storage and retrieval works~\cite{Dong}.

For the experimental demonstration, optical fields in a whispering gallery
mode (WGM) of a silica microsphere ($\kappa/2\pi\sim 30$MHz) with a diameter of $33\mu $\textrm{m}
were coupled to the (1, 2) radial-breathing mechanical mode ($\omega
_{m}/2\pi \thicksim 94$\textrm{MHz}, $\gamma _{m}/2\pi \thicksim 20$\textrm{%
kHz} and $Q\sim1.3\times10^7$) of the microsphere. The WGM was excited through the evanescent field
of a tapered optical fiber. Both the silica microsphere and the tapered
fiber were held in a clean room environment in order to avoid degradation of
optical Q factors. In Figure 3(b-c), we show the WGM transmission resonance
used in our experiment and the displacement power spectrum of the (1 ,2)
radial breathing mode of the microsphere \cite{park}. A combination of
acoustic-optic modulator (AOM) and electro-optic modulators (EOMs) were used
to generate optical pulses with the desired duration, timing, and
frequencies, as shown in Fig.3 (a). The driving and the locking pulses came
from a single-frequency tunable diode laser (Toptica DLPRO 780) with $%
\lambda \thicksim 780$ \textrm{nm} and with its frequency locked to the red
sideband of a given WGM resonance using the Pound-Drever-Hall technique.
The signal pulses were drived from the blue sideband
generated by passing the driving pulses through EOM$_{0}$.%
For the experimental results reported here, both the driving and the probe
pulses were square shaped, with the same timing and with duration of $4\mu $%
\textrm{s}. Heterodyne detection was used for the measurement of
the optical emission from the microsphere near the WGM resonance, with the
driving laser pulse serving as the local oscillator \cite{Dong}. A gated
detection scheme was also used with a gate duration of $1\mu $\textrm{s}.
The timing of the gate (see grey area in Fig.\ref{fig1}c) determines the
effective duration, $\tau_{2}$, of the second pulse pair involved in the RI.

\begin{figure}[tbph]
\includegraphics[width=0.48\textwidth]{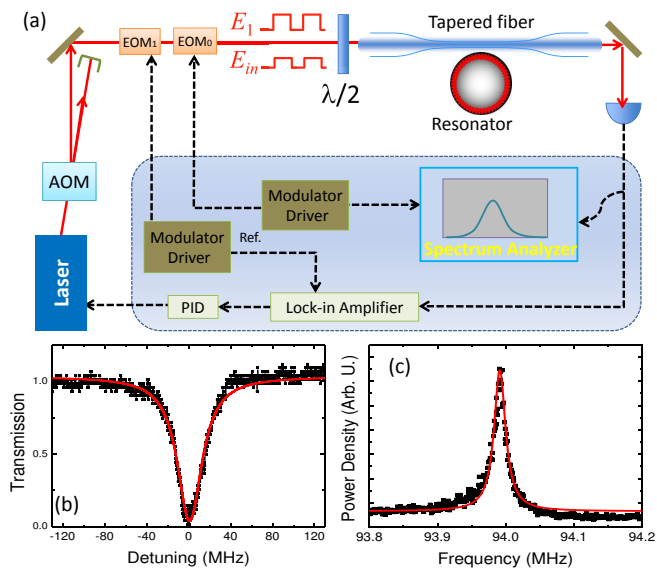}
\caption{{(a) Schematic of the experimental setup for the Ramsey
Interferometry, with solid lines representing the optical paths and dashed
lines representing the electrical connections. (b) Transmission resonance
for the WGM at $\sim 780$nm. (c) Displacement power spectrum of the (1, 2)
radial breathing mode, obtained from the same sample. The solid red lines
show a Lorentzian fitting. }}
\label{fig4}
\end{figure}

\begin{figure}[tbph]
\includegraphics[width=0.48\textwidth]{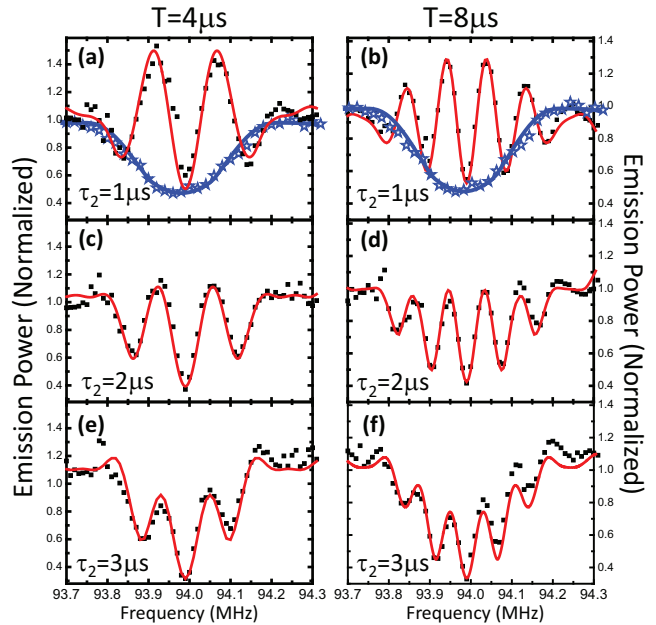}
\caption{Heterodyne detected probe intensity as a function of the detuning
between the probe and driving lasers.{\ The blue dots in (a) and (b) are
obtained with the detection gate positioned within the first pulse pair with
the dealy time $t^{\prime }=3\protect\mu $s. The black dots in (a-f) are
obtained with the detection gate positioned within the second pulse pair
with $T$ and $\protect\tau_{2}$ indicated in the figure. }The incident
driving power used is $3.4$\textrm{mW}. The solid lines are results of
theoretical calculations discussed in Sec. III. }
\label{fig3}
\end{figure}

Figures~\ref{fig3} show the heterodyne detected probe intensity as a
function of the detuning between the probe and driving laser obtained with
different durations $\tau_{2}$. The separation time, T,
between the first and second pulse pairs is set to $4\mu $\textrm{s} in the
left column and $8\mu $\textrm{s} in the right column. The distinct spectral
oscillations observed in these experiments demonstrate the Ramsey fringes
for the OMS. As a reference, we also show in Figs.~\ref{fig3}a and \ref{fig3}%
b the experimental results (solid circles) obtained in the absence of the
second pulse pair. Experimentally, these were obtained with the detection
gate positioned within the duration of the first pulse pair, as indicated in
Fig.~\ref{fig1}c. The spectral dip observed in this case arises from \textit{the transient}
optomechanically induced transparency (OMIT)~\cite{EIT1,EIT2,EIT3}.

\section{Theoretical model}
For a detailed theoretical analysis of optomechanical Ramsey fringes and for
a direct comparison between experiment and theory, we use the optomechanical
Hamiltonian written as~\cite{review1,review2}%
\begin{align}
H& =\hbar (\omega _{c}-\omega _{l})a^{\dag }a+\hbar \omega _{m}b^{\dag
}b-\hbar ga^{\dag }a(b^{\dag }+b)  \notag  \label{01} \\
& \quad +i\hbar \mathcal{E}_{l}(t)(a^{\dag }-a)+i\hbar \lbrack \mathcal{E}%
_{p}(t)a^{\dag }\mathrm{e}^{-i(\omega _{p}-\omega _{l})t}+c.c.],
\end{align}%
where we denote the optical and mechanical modes using annihilation
operators $a$ and $b$, respectively. The first line in Eq. 1 contains the
terms of the optical and mechanical unperturbed Hamiltonian and the
interaction between them; and the terms in the second line describe
interaction with the driving laser and the probe laser. The linear coupling
rate g is given by $(\omega _{c}/L)\sqrt{\hbar /(m\omega _{m})}$\ with $L$
being the cavity length and m being the effective mechanical mass. The total
decay rate of the cavity field includes the intrinsic decay rate $\kappa
_{i} $, the rate at which the optical energy losses to the environment, and the external decay rate $\kappa _{e}$ which is the loss rate associated with the waveguide-resonator interface. The driving and probe laser amplitudes are related to their powers $\mathcal{P}_{i}$ by $\mathcal{E}_{i}=\sqrt{\kappa \mathcal{P}_{i}/(\hbar \omega _{i})}$. We assume that the
coupling laser is much stronger than the probe laser, $\mathcal{E}_{l}\gg
\mathcal{E}_{p}$.

We follow the standard procedure to solve the problem by expanding the mean
values of equations to the first order in $\mathcal{E}_{p}$ using $a=\alpha
_{0}+\alpha \mathrm{e}^{-i(\omega _{p}-\omega _{l})t}$ and $b=\beta
_{0}+\beta \mathrm{e}^{-i(\omega _{p}-\omega _{l})t}$, and get $\alpha
_{0}(t)=\frac{\mathcal{E}_{l}(t)}{\kappa /2+i\Delta }$, $\beta _{0}(t)=\frac{%
g}{\omega _{m}-i\gamma _{m}/2}|\alpha _{0}|^{2}$ with $\Delta =\omega
_{c}-\omega _{l}-2|\alpha _{0}|^{2}g^{2}/\omega _{m}$ representing the
effective detuning between the cavity and the driving field frequencies. The
driving-enhanced coupling rate $G(t)=|\alpha (t)|g$ is solely controlled by
the coupling laser amplitude and we denote $G(t)=G$ during the pulses and $%
G(t)=0$ otherwise. In the equations for the first order amplitudes $\alpha $
and $\beta $, we make the rotating-wave approximation by eliminating the
non-resonant fast oscillating terms $\mathrm{e}^{-2i(\omega _{p}-\omega
_{l})t}$ and $\mathrm{e}^{2i(\omega _{p}-\omega _{l})t}$. Physically, we
ignore the terms corresponding to the processes of creating or eliminating
both a photon and a phonon simultaneously. The simplified equations for the
amplitudes $\alpha $ and $\beta $ are given by
\begin{equation}
\begin{aligned}
& \dot{\alpha} = -ix\alpha - G(t)\beta - (\kappa/2) \alpha + \mathcal{E}_p(t), \\
& \dot{\beta} = -iy\beta + G(t)\alpha - (\gamma_m/2)\beta.
\end{aligned}  \label{03}
\end{equation}%
The output optical field at frequency $%
\omega _{p}$ at any time $t$ can be derived from the input-output relation $%
\mathcal{E}_{o}(t)=\kappa _{e}\alpha (t)-\mathcal{E}_{p}(t)$. In our study,
we work in the critical coupling regime with the internal and external
coupling rates equal, i.e. $\kappa _{e}=\kappa _{i}=\kappa /2$. The $\kappa$ gives the linewidth of the transmission around the WGM which for the data of Fig.~\ref{fig2} is $2\pi\times30$MHZ.  We do
heterodyne detection to get the intracavity field $\kappa _{e}\alpha (t)=%
\mathcal{E}_{o}(t)+\mathcal{E}_{p}(t)$.

We now study the mechanical field $\beta _{R}$ and optical field $\alpha
_{R} $ after application of the two separated pulses. We give the approximate
expressions which are normalized to $\mathcal{E}_{p}$ by solving Eqs.(\ref%
{03}):
\begin{equation}\label{04}
\begin{aligned}
& \kappa _{e}\beta _{R}\approx G\left[ \frac{\mathrm{e}^{-(iy+\Gamma )\tau
_{1}}-1}{iy+\Gamma }\mathrm{e}^{-i\phi-\mu }+\frac{\mathrm{e}^{-(iy+\Gamma )\tau
_{2}}-1}{iy+\Gamma }\right] \mathcal{E}_{p}.   \\
& \kappa _{e}\alpha _{R}\approx (2G\beta _{R}+1)\mathcal{E}_{p},  \\
& \phi = y(\tau_{2}+T), \quad \mu= (\frac{\gamma _{m}}{2}T + \Gamma \tau_{2}), \quad y=\omega_l+\omega_m-\omega_p,
\end{aligned}
\end{equation}%
where $\Gamma = 2G^{2}/\kappa +\gamma _{m}/2$ is the photon-phonon transfer
rate in OMS and $\phi $ denotes the accumulated phase which leads to the
Ramsey fringes. The parameter $\mu$ describes the decay of the signal, which causes the loss of visibility in the Ramsey fringes. The two terms in $\mu$ are corresponding to the mechanical decay in the free evolution time $T$ and in the optomechanical interaction during the second pulse $\tau_2$.

We start by analyzing the mechanical mode given in Eq.(\ref{04}). It
contains two terms with the same form, each of which describes the phonon
excitation due to the optomechanical interaction when the corresponding
pulse $\tau_{i}$ ($i=1,2$.) is on. However, there is a phase factor $%
\mathrm{e}^{-i\phi }$ multiplied to the first term. As $\phi $ changes, the
two terms in Eq.(\ref{04}) either interfere constructively or destructively
leading to the Ramsey fringes. The phase $\phi=y(T+\tau_{2})$, i.e. the product of the frequency detuning and the evolution time, determines the fringe period. Note that the fringe period also depends on the parameter $\Gamma$. The period is exactly $2\pi /(T+\tau_{2})$ only when $\Gamma=0$ and it gets lower when $\Gamma$ increases. The fringes get sharper as one increases $T+\tau_{2}$. However, a longer $\tau_{2}$ can result in a decay of the signal, which can be seen from the parameter $\mu= \Gamma \tau_{2}+\frac{\gamma _{m}}{2}T$. Considering $\Gamma >\gamma _{m}$, one should reasonably choose $\tau_{2}<1/\Gamma $.
The numerator of each term $\mathrm{e}^{-(iy+\Gamma )\tau_{i}}-1\sim 0$ for
a short $\tau_{i}$, and it increases along with $\tau_{i}$. This justifies
that the phonon excitation is only prominent when $\tau_{i}$ is large,
which sets the characteristic time of phonon excitation in OMS, i.e. $\Gamma
\tau_{i}>1$. The electromagnetically induced transparency (EIT) occurs when
$\Gamma \tau_{i}\gg 1$. In the optomechanical RI, a large $\tau_1$ enhances the Ramsey fringes contrast, although the fringes can still be seen at a shorter $\tau_1$. Therefore, the conditions for the Ramsey fringes are $\tau_{1}\Gamma\gtrsim1$, $\tau_{2}\Gamma<1$, and $T\gamma_{m}\ll1$.

The optical field expressed in Eqs.(\ref{04}) exhibits the same interference
fringes as in the mechanical mode. This is important as the measurement of the output optical field becomes a direct probe of the Ramsey fringes in the mechanical system.  For a direct comparison with the experiments, over a wide range of parameters we show the results of the theoretical calculations as solid curves in Fig.~\ref{fig3}. The parameters used include $\kappa /2\pi =30$%
MHz, $G/2\pi =0.58$MHz, $\gamma _{m}/2\pi =20$kHz, and the corresponding characteristic time $1/\Gamma=4.9\mu$s. As shown in Fig.~\ref{fig3}, the spectral position of the central Ramsey fringe overlaps exactly
with the center of the OMIT dip and does not depend on either $T$ or $\tau
_{2}$. More importantly, the Ramsey fringes exhibit a period that is much
smaller than the linewidth of the OMIT dip. In (a) with $T=4\mu s$ and $\tau
_{2}=1\mu s$, the Ramsey fringe period is $160$\textrm{kHz}. As $T+\tau_{2}$
increase from $5\mu s$ in (a) to $11\mu s$ in (f), the fringe period
decreases from $\thicksim 160$\textrm{kHz} to $\thicksim 80$\textrm{kHz}. Overall, there is an
excellent agreement between the theory (curves) and experiment (dots).
The visibility of the Ramsey fringes is primarily determined by $\tau_{2}$.
Fig.~\ref{fig3}. reveals the loss of fringe visibility with increasing $\tau_{2}$. We note that for comparison with the experiments, we use directly Eqs.(\ref{03}). It is only for understanding the physical behavior we used approximate Eqs.(\ref{04})

\begin{figure}[tbph]
\includegraphics[width=0.48\textwidth]{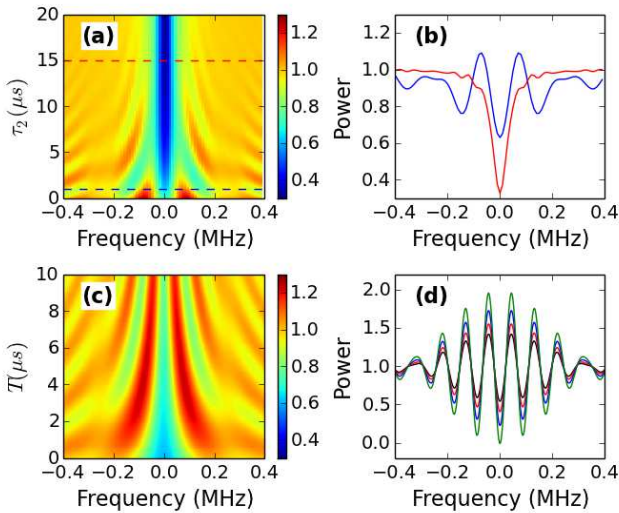}
\caption{(a) Simulation as a function of $\tau_2$ with fixed $\tau_1=4\mu$s, $T = 4\mu$s, $\kappa/2\pi=30$MHz, $G/2\pi = 0.58$MHz, $\gamma_m/2\pi = 20$kHz. (b) The spectra with $\tau_2=1\mu$s (blue curve) and $\tau_2=15\mu$s (red curve), as dashed line shown in (a). (c) The Ramsey fringe with different delay time $T$. Other parameters are same as (a) except $\tau_2=0.1\mu$s. (d) The Ramsey fringes with $T=10\mu$s and different $\gamma_m/2\pi=10, 20, 30, 40k$Hz for green, blue, red and black curves, respectively. Other parameters are same as in (c). }
\label{fig5}
\end{figure}
In order to appreciate the versatility of the Ramsey fringes in OMS, we show in Fig.~\ref{fig5} additional results of simulations under a range of parameters.  As shown in
Fig.~\ref{fig5}a, the interference fringes decrease with the longer
$\tau_2$, which means no coherent phonons for interference. This is due to the decay term $\mu$ in Eq.(\ref{04}). As noted there we need $\Gamma\tau_2$ to be small.  In Fig.~\ref{fig5}b we can see that, when $\tau_2$ is short, the spectrum shows interference; but with the long enough $\tau_2$, the fringes in the spectrum cannot be seen and the spectrum reduces to a steady-state result. With
other parameters fixed, the increase in time $T$ leads to a deceasing
in the Ramsey fringe period. However, after long
enough time, the Ramsey fringes disappear because of
the damping of phonons, as shown in Fig.~\ref{fig5}c. Therefore, we should choose the $\gamma_m$ as small as possible during the
experiment for observing Ramsey fringes. This is demonstrated more clearly in Fig.~\ref{fig5}d, which shows the visibility of the Ramsey fringes with different $\gamma_m$.

\section{Conclusion}
In conclusion we have demonstrated how the high resolution Ramsey method of
separated oscillatory fields can be adopted to study coherences in a
macroscopic system like a nanomechanical oscillator. We have presented the
underlying theory and the experimental demonstration using silica
microresonators. The method is quite versatile and can be adopted to
different types of mechanical resonators and electromechanical resonators.
More complex applications can include the study of the dynamical interaction
between the mechanical oscillators. Future work may also include the
demonstration of the Ramsey fringes using excitations at the single photon
level which would imply excitation of a mechanical oscillator at the single
phonon level needless to say that achieving quantum regime experimentally would require at least the coherent fields at the single photon level as well as cooling to temperatures such that the mean phonon number is less than $1$. The Ramsey method is also expected to be useful in producing
time-bin entanglement involving a phonon and a photon.

K.Q. and C.H.D. contributed equally to this work. C.H.D. was supported by
the National Basic Research Program of China (Grant No.2011CB921200), the
Knowledge Innovation Project of Chinese Academy of Sciences (Grant
No.60921091), the National Natural Science Foundation of China (Grant
No.61308079), the Fundamental Research Funds for the Central Universities.
H.W. acknowledges support from NSF under grant No. 1205544.

\end{document}